\documentstyle[12pt]{article}
\author{Afraimovich~E.~L.\\
Institute of Solar-Terrestrial Physics SD RAS,\\
p.~o.~box~4026, Irkutsk, 664033, Russia\\
fax: +7 3952 462557; e-mail:~afra@iszf.irk.ru}
\title{The GPS global detection of the ionospheric response to
solar flares}
\date{}
\begin{document}
\maketitle
\begin{abstract}
This author suggests the concept of a new technology for global
detection (GLOBDET) of atmospheric disturbances of natural and
technogenic origin, on the basis of phase measurements of the
total electron content (TEC) in the ionosphere using an
international GPS network. Temporal dependencies of TEC are
obtained for a set of spaced receivers of the GPS network
simultaneously for the entire set of "visible" (over a given time
interval) GPS satellites (up to 5-10 satellites). These series
are subjected to filtering in the selected range of oscillation
periods using algorithms for spatio-temporal analysis of signals
of non-equidistant GPS phased antenna arrays which are
adequate to the detected disturbance. An analysis is made of the
possibilities of using the GLOBDET when detecting the ionospheric
response of solar flares. In this case it is best to make the
coherent summation of the filtered series of TEC. A powerful
impulsive flares of July 29, 1999 and December 28, 1999 were chosen
to illustrate the practical implementation of the proposed
method.
\end{abstract}
\section{Introduction}
\label{GLOB-Sect1}

The advent and evolution of a Global Positioning System, GPS, and
also the creation on its basis of widely branched networks of GPS
stations (at least 700 sites at the May of 2000, the data from which are
placed on the INTERNET) opened up a new era in remote ionospheric
sensing \cite{Klo97}. Furthermore, there exist also powerful regional
networks such as the Geographical Survey Institute network in Japan
\cite{Sai98} consisting of up to 1000 receivers. High-precision
measurements of the group and phase delay along the line-of-sight
(LOS) between the receiver on the ground and transmitters on the
GPS system satellites covering the reception zone are made using
two-frequency multichannel receivers of the GPS system at almost
any point of the globe and at any time simultaneously at two
coherently coupled frequencies $f_1=1575{.}42$ MHz and
$f_2=1227{.}60$ MHz.

The sensitivity of phase measurements in the GPS system is
sufficient for detecting irregularities with an amplitude of up
to $10^{-3}$--$10^{-4}$ of the diurnal TEC variation. This makes
it possible to formulate the problem of detecting ionospheric
disturbances from different sources of artificial and natural
origins. Anthropogenic sources of ionospheric disturbances
include nuclear events, chemical explosions, and rocket launches.
Among natural sources are solar eclipses, solar flares,
earthquakes, volcanoes, heavy thunderstorms, and auroral heating.
Studies of these phenomena are of great importance both for a
scientific understanding of their genesis and propagation in the
atmosphere and as a tool for detecting such sources in remote
regions; for instance, the GLONASS system provides more effective
coverage at high latitudes \cite{Klo97}.

Recently some authors embarked actively on the development of
detection tools for the ionospheric response of powerful
earthquakes \cite{Cal95}, rocket launches \cite{Cal96}, and
industrial surface explosions \cite{Fit97}; \cite{Cal98}.
Subsequetly, the GPS data began to be used in the context of the
spaced-receiver method using three GPS stations to determine the
parameters of the full wave vector of traveling ionospheric
disturbances under quiet and disturbed geomagnetic conditions
\cite{Afr98}; \cite{Afr2000}.

The limitations of the spaced-receiver method with a minimum
necessary number of receivers (three) include their low
sensitivity and inadequate spatial selectivity, and this gives no
way of exploiting the potential of a global GPS system consisting
of hundreds of GPS stations. Next in turn is the setting of a
more general problem, namely, that of developing processing
techniques for the data from the GPS and GLONASS systems, based
on the latest achievements in the field of a comprehensive
spatio-temporal processing of signals, with due regard for the
special features of ionospheric disturbances, however.

The objective of this paper is to develop the concept of a global
detector (GLOBDET) of atmosperic disturbances of natural and
technogenic origins, based on phase measurements of total electron
content (TEC) in the ionosphere using the international network of
two-frequency multichannel receivers of the GPS system - section
~\ref{GLOB-Sect2}. Section ~\ref{GLOB-Sect3} discusses the
possibilities of using the GLOBDET method in detecting ionospheric
effects of solar flares. Basic characteristics of such a detector
(sensitivity and time resolution) are examined in Section
~\ref{GLOB-Sect4}. The practical implementation of the method is
illustrated by analyzing a global ionospheric response to a powerful
solar flares of July 29, 1999 and December 28, 1999 (Sections
~\ref{GLOB-Sect5} and ~\ref{GLOB-Sect6}). Section
~\ref{GLOB-Sect7} discusses the findings and explores the
possibilities of utilizing a global GPS network in detecting other kinds
of atmospheric disturbances of natural and technogenic origins.

\section{Using the data from a global GPS network in the context
of the ideology of a phased antenna array}
\label{GLOB-Sect2}

Phased antenna arrays (FAA) are extensively used in radiolocation,
sonar systems, and in processing seismic signals as systems
featuring high sensitivity in signal detection and high spatial
selectivity which is necessary for localizing signal sources.
There exist quite varied schemes for hardware and software
realization of FAA \cite{Col85}. In general case the treatment of
signals imply multiplying a complex signal $\tilde A(t)_i$ of
each of the $i$--spaced FAA elements by a complex factor $\tilde
K_i$ with the subsequent coherent summation of the resulting
products

\begin{equation}
\label{GLOB-eq-01}
\tilde S = \sum^N_{i=1} \tilde A(t)_i  \tilde K_i
\end{equation}

where $\tilde S$ is the result of the coherent summation; $i$ is
the FAA element number; and $i$=1, 2, ...$N$.

By specifying different values of $\tilde K_i$, which in each
particular case depend on the conditions of the problem being
solved, it is possible to specify parameters determining the
sensitivity and selectivity of FAA.

This author suggests the concept of a global detector (GLOBDET)
of atmospheric disturbances of natural and technogenic origins,
based on processing the GPS data in the FAA configuration. This
concept implies that time dependencies of TEC are obtained for a
set of spaced receivers of the GPS network simultaneously for all
GPS satellites "visible" during a given time interval. These
series are subjected to filtering over a selected range of
oscillation periods in order to eliminate slow variations caused
by the orbital motion of the satellites and by the diurnal
variation of TEC in the ionosphere. Next, these series are
processed in the nonequidistant FAA configuration using them in
~(\ref{GLOB-eq-01}) or in other algorithms of PAA as complex
signals $\tilde A(t)_i$. It is also assumed that equivalent
partial antennas of such an array are located at subionospheric
points. The coordinates of these points are determined for the
selected height $h_{max}$ of the $F2$-layer maximum by means of
standard (in the GPS system) calculations of the azimuth $\alpha$
and elevation $\theta$ of the LOS between the
receiver and the satellite. $\alpha$ and $\theta$ are reckoned
from the northward direction and from the ground, respectively.

The GPS technology provides the means of estimating TEC
variations on the basis of phase measurements of TEC $I$ in each
of the spaced two-frequency GPS receivers using the formula
\cite{Hof92}; \cite{Cal96}:

\begin{equation}
\label{GLOB-eq-02}
I=\frac{1}{40{.}308}\frac{f^2_1f^2_2}{f^2_1-f^2_2}
                           [(L_1\lambda_1-L_2\lambda_2)+const+nL]
\end{equation}

where $L_1\lambda_1$ and $L_2\lambda_2$~ are phase path
increments of the radio signal, caused by the phase delay in the
ionosphere (m); $L_1$, $L_2$~ is the number of full phase
rotations, and $\lambda_1$, and $\lambda_2$, are the wavelengths
(m) for the frequencies $f_1$ and $f_2$, respectively; $const$~
is some unknown initial phase path (m); and $nL$~ is the error in
determination of the phase path (m).

The TEC unit, $TECU$, which is equal to $10^{16}$ m${}^{-2}$ and is commonly
accepted in the literature, will be used throughout the text.

\section{Using the GLOBDET technology in detecting the ionospheric
response to solar flares}
\label{GLOB-Sect3}

The enhancement of X-ray and ultraviolet radiation intensity
that is observed during chromospheric flares on the Sun
immediately causes an increase in electron density in the
ionosphere. These density variatiuons are different for
different altitudes and are collectively called Sudden
Ionospheric Disturbances (SID). SID observations provide a key
means for ground-based detection of solar flares along with
optical observations of flares and solar radio burst
observations. Much research is devoted to SID studies, among
them a number of thorough reviews and monographs \cite{Mit74}.

SID data for the $F$-region acquired by different radio probing
methods were used repeatedly to estimate time variations in the
X-ray and extreme ultraviolet (EUV) spectral regions and in
relative measurements of fluxes in different wavelength ranges
\cite{Don69}; \cite{Tho71}; \cite{Men74}. The main body of SID
data for the Earth's upper atmosphere was obtained in earlier
detections of Sudden Frequency Deviations (SFD) of the
$F$-region-reflected radio signal in the HF range \cite{Dav69};
\cite{Don69}.

A further, highly informative, technique is the method of incoherent
scatter (IS) \cite{Tho71}. However, the practical implementation of
the IS method requires very sophisticated, expensive equipment. An
added limitation is inadequate time resolution. Since the relaxation
time of electron density in the $E$- and $F1$-regions is also less than
5-10 min, most IS measurements lack time resolution needed for the
study of inospheric effects of flares.

The effect of solar flares on the ionospheric F-region is also
manifested as a Sudden Increase of Total Electron Content (SITEC)
which was measured previously using continuously operating radio
beacons on geostationary satellites \cite{Men74}. A serious
limitation of methods based on analyzing VHF signals from
geostationary satellites is their small and ever increasing with
the time number and the nonuniform distribution in longitude.

Consequently, none of the above-mentioned existing methods can
serve as an effective basis for the radio detection system to
provide a continuous, global SID monitoring with adequate
space-time resolution. Furthermore, the creation of these
facilities requires developing special-purpose equipment,
including powerful radio transmitters contaminating the radio
environment. It is also significant that when using the existing
methods, the inadequate spatial aperture gives no way of deducing
the possible spatial inhomogeneity of the X-ray and EUV flux.

According to the above concept, a global GPS network can be
successfully used as a global detector of the ionospheric
response to solar flares. A physical groundwork for the method is
formed by the effect of fast change in electron density in the
Earth's ionosphere at the time of a flare simultaneously on the
entire sunlit surface.

Essentially, the method implies using appropriate filtering and a
coherent processing of TEC variations in the ionosphere
simultaneously for the entire set of "visible" (during a given
time interval) GPS satellites (as many as 5-10 satellites) at all
global GPS network stations used in the analysis. In detecting
solar flares, the ionospheric response is virtually simultaneous
for all stations on the dayside of the globe within the time
resolution range of the GPS receivers (from 30 s to 0.1 s).
Therefore, the coherent addition of individual realizations
$\tilde A(t)_i$ does not require a partial phase shift (the
complex part of the term $\tilde K_i$ is zero), and the entire
procedure reduces to a simple addition

\begin{equation}
\label{GLOB-eq-03}
\tilde S = \sum^N_{i=1} \tilde A(t)_i K_i
\end{equation}

where $\tilde A(t)_i$ represents filtered TEC variations, and
$K_i$ is the amplitude weighting factor determined by the
geometry of the $i$ -beam (LOS) to the satellite. To a first
approximation, this factor is \cite{Klo86}.

\begin{equation}
\label{GLOB-eq-04}
K_i = cos \left[arcsin\left(\frac{R_z}{R_z + h_{max}}cos\theta_i\right)
\right]
\end{equation}

where $R_z$ is the Earth's radius; $h_{max}=300$ is
the height of the $F2$--layer maximum.

Reconstructing the absolute value of the ionospheric response to
the solar flare requires a more accurate (than used in this
paper) conversion of the "oblique" TEC value to a "vertical" one,
especially at low values of elevations of the beam to the
satellite. To do this, it is necessary not only to eliminate, for
this beam to the satellite, the ambiguity of the determination of
the absolute TEC value which arises when only phase measurements
are used in the GPS system. The response can only be estimated
reliable, with the inclusion the spatially inhomogeneous
ionosphere, by using all beams to the satellite, and by applying
adequate methods of spatial interpolation. This problem is
considered in a series of publications (for example, \cite{Man98})
and is beyond the scope of this paper.

\section{Characteristics of a global detector}
\label{GLOB-Sect4}

The detection sensitivity is determined by the ability to detect
typical signals of the ionospheric response to a solar flare
(leading edge duration, period, form, length) at the level of TEC
background fluctuations. Ionospheric irregularities are
characterized by a power spectrum, so that background
fluctuations will always be distinguished in the frequency range
of interest. However, background fluctuations are not correlated
in the case of beams to the satellite spaced by an amount
exceeding the typical irregularity size.

With a typical length of X-ray bursts and EUV emission of solar
flares of about 5-10 min, the corresponding ionization
irregularity size does normally not exceed 30-50 km; hence the
condition of a statistical independence of TEC fluctuations at
spaced beams is almost always satisfied. Therefore, coherent
summation of responses to a flare on a set of beams spaced
thoughout the dayside of the globe permits the solar flare effect
to be detected even when the response amplitude on partial beams
is markedly smaller than the noise level (background
fluctuations). The proposed procedure of coherent accumulation is
essentially equivalent to the operation of coincidence schemes
which are extensively used in X-ray and gamma-ray telescopes.

If the SID response and background fluctuations, respectively,
are considered to be the signal and noise, then as a consequence
of a statistical independence of background fluctuations the
signal/noise ratio when detecting the flare effect is increased
through a coherent processing by at least a factor of $\sqrt{N}$,
where $N$ is the number of LOS.

The solar flares of July 29, 1999 and Decevber 28, 1999 were used to
illustrate the performance of the proposed method. Fig.1a presents the
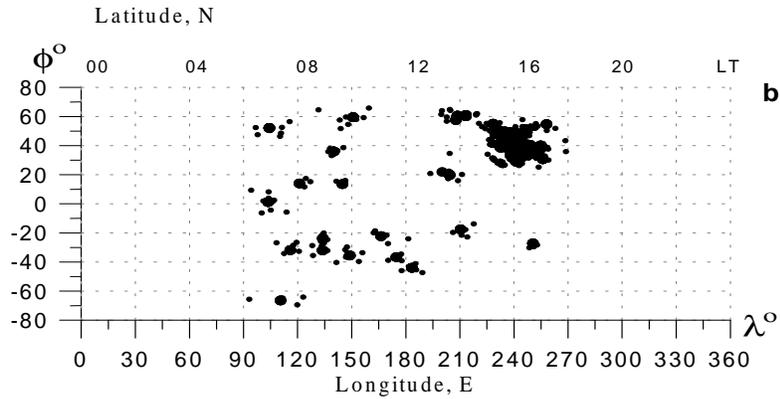
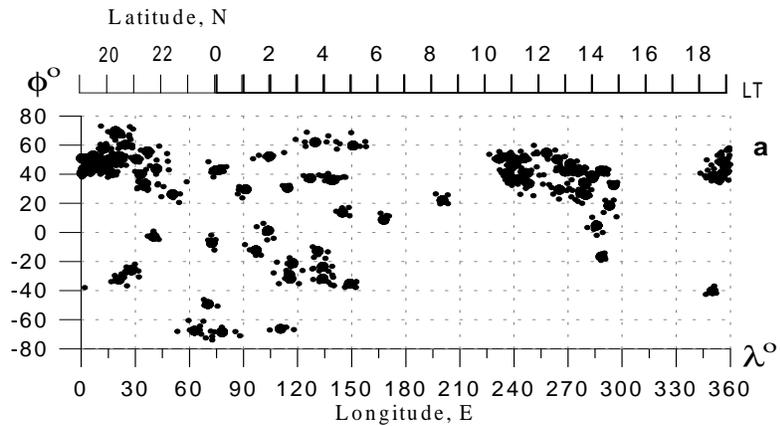
geometry of a global GPS array used in this paper to analyze the
effects of the July 29, 1999 flare (105 stations). Heavy dots correspond
to the location of the GPS stations. The upper scale indicate the local
time, LT, corresponding to 19:00 UT. The coordinates of the stations
are not given here for reasons of space.

As is evident from Fig. 1a, the set of stations which we chose out
of the global GPS network available to us, covers rather densely
North America and Europe, but provides much worse coverage of the
Asian part of the territory used in the analysis. The number of
GPS stations in the Pacific and Atlantic regions is even fewer.
However, coverage of the territory with partial beams to the
satellite for the limitation on elevations $\theta>10^\circ$~,
which we have selected, is substantially wider. Dots in Fig. 1a
mark the coordinates of subinospheric points for the height of
the $F2$--layer maximum $h_{max}=300$ km for all visible
satellites at 19:00 UT for each GPS station. A total number of
beams (and subionospheric points) used in this paper to analyze
the July 29, 1999 flare is 622.

Fig.1b presents the geometry of a global GPS array used in this paper
to analyze the effects of the December 28, 1999 solar flare, (230
stations; only for the dayside). A total number of beams (and
subionospheric points) is 1200.

Such coverage of the terrestrial surface makes it possible to
solve the problem of detecting time-coincident events with
spatial resolution (coherent accumulation) two orders of
magnitude higher, as a minimum, than could be achieved in SFD
detection on oblique HF paths. For simultaneous events in the
western hemisphere, the correspoding today's number of stations
and beams can be as many as 400 and 2000--3000, respectively.

It should be noted that because of the relatively low satellite
orbit inclinations, the GPS network (and to a lesser degree
GLONASS) provides poor coverage of the Earth's surface near the
poles. However, TEC measurements in the polar regions are
ineffective with respect to the detection of the ionospheric
response to a solar flare because the amplitude of background
fluctuations in this case is much higher when compared with the
mid-latitude ionosphere. This is partiocularly true of
geomagnetic disturbance periods. For the same reason, equatorial
stations should also be excluded from a coherent processing.

If the Earth's ionosphere is regarded as the filling of some
global detector of X-ray and EUV emissions, then it is possible
to estimate a huge volume $V_{det}$ of the potential sensitivity
region of such a detector. This volume is equal to one-half the
difference of the volumes of spheres with the radii
$R_{z}+H_{max}$ and $R_{z}+H_{min}$

\begin{equation}
\label{GLOB-eq-05}
V_{det}=2 \pi [(R_{z}+H_{max})^3-(R_{z}+H_{min})^3]/3
\end{equation}

where $H_{min}$ and $H_{max}$ are the upper and lower boundaries
of the absorbing layer for a given part of the flare emission
spectrum. For the EUV range,
$H_{min}=100$ km, and $H_{max}=200$ km \cite{Mit74}, which gives
the volume $V_{det}$ of order $2.65\,10^{19}$ m${}^{3}$. For the
X-ray part of the spectrum, $H_{min}=60$ km, and $H_{max}=80$ km
\cite{Mit74}, and the corresponding volume is $0.523\,10^{19}$
m${}^{3}$.

The actual sensitivity is in fact determined by the number of beams
penetrating the region of potential sensitivity. Furthermore, solar
flare-induced TEC perturbations constitute but a small part of TEC
including the height range up to 1000 km. There are methods to
estimate the height in the ionosphere, but they do not come within the
province of this paper.

On the one hand, GLOBDET time resolution is limited by technical
capabilities of the GPS system. Essentially, data with a time
resolution of about 30 s are currently available on the INTERNET,
which is insufficient for a detailed analysis of the fine
structure of the SID time dependence. Yet this limitations seems
to be transient since current multichannel two-frequency GPS
receivers can operate with a time resolution of up to 0.1 s.

On the other hand, time resolution is determined by time
constants of ionization and recombination processes in the
ionosphere at a given height \cite{Don69}, \cite{Mit74}; these
parameters can be taken into acount when processing the data.

\section{Ionospheric response to the solar flare of July 29, 1999}
\label{GLOB-Sect5}

A powerful impulsive flare of July 29, 1999 was chosen to
illustrate the practical implementation of the proposed method.
The thick line in Fig. 2d shows the time dependence of the X-ray
emission intensity $R(t)$ of this flare as measured by the
X-telescope BATSE on satellite CGRO in 25--50 keV range. The
envelope of the dependence $R(t)$ represents a bell-shaped
pulse of about 3-min duration with a maximum corresponding to
19:34 UT, or to about 12:00 LT in the U.S. West (Fig. 1a). This
time interval is characterized by a low level of geomagnetic
disturbance (within -10 nT), which simplified greatly the SID
detection problem.

We now describe briefly the sequence of GPS data processing
procedures. Primary data include series of "oblique" values of
TEC $I(t)$, as well as the corresponding series of elevations
$\theta(t)$ and azimuths $\alpha(t)$ along LOS to the
satellite calculated using our developed CONVTEC program which
converts the GPS system standard RINEX-files on the INTERNET
\cite{Gur93}. The determination of SID characteristics involves
selecting continuous series of $I(t)$ measurements of at least a
one-hour interval in length, which includes the time of the
flare. Series of elevations $\theta(t)$ and azimuths
$\alpha(t)$ of the beam to the satellite are used to determine
the coordinates of subionospheric points. In the case under
consideration, all results were obtained for elevations
$\theta(t)$ larger than $10^\circ$.

Fig.~2a presents typical time dependencies of an "vertical" TEC
$I(t)$ for the PRN03 satellite at the CME1 station
($40.4^\circ$N; $235.6^\circ$E) on July 29, 1999 (thick line) and
for PRN21 at the CEDA station ($40.7^\circ$N; $247.1^\circ$E --
thin line). It is apparent from Fig.~2a that in the presence of
slow TEC variations, the SID-induced short-lasting sudden
increase in TEC is clearly distinguished in the form of a "step"
as large as 0.4 $TECU$.

For the same series, similar lines in panel b. show variations of
the time derivative of TEC $dI(t)/dt$ with the linear trend
removed and with a smoothing with the 5-min time window. The TEC
time derivative is invoked because it reflects electron density
variations which are proportional to the X-ray or EUV flux
\cite{Mit74}.

The coherent summation of $dI(t)/dt_i$ realizations was made by the
formula

\begin{equation}
\label{GLOB-eq-06}
S(t) = \sum^N_{i=1} dI(t)/dt_i K_i
\end{equation}

The (normalized to $N$) result of a coherent summation of the
$S(t)$-series for all beams and GPS stations located mainly on
the dayside is presented in panel c (thick line). A comparison of the
resulting coherent sum of $S(t)$ with the time dependence
$dI(t)/dt$ for individual beams presented in panel b.
confirms the effect of a substantial increase of the signal/noise
ratio caused by a coherent processing.

It is interesting to compare, for the same time interval, the results from
a coherent summation for the dayside and nightside. The r.m.s. of the
coherent sum of $S(t)$ in panel c. for the nightside (thin line) is an
order of magnitude (as a minimum) smaller than the SID response
amplitude.

A comparison of the coherent sum of $S(t)$ with the time
dependence of the X-ray emission intensity $R(t)$ of the July
29, 1999 flare, based on the data from the X-telescope BATSE on
satellite CGRO data in 25--50 keV range (panel d), shows their high
correlation and an almost total coincidence of the form of $S(t)$
with the $R(t)$ pulse envelope.

It should be noted, however, that generally TEC variations $S(t)$
are 60 s ahead of $R(t)$. It is pointed out in \cite{Mit74}
that EUV emission is mostly responsible for SID in the
$F$-region, TEC variations are also well correlated with X-ray
flares. This is also confirmed by simultaneous measurements of
flare emission characteristics in the X-ray and EUV ranges by the
Solar Maximum Mission spacecraft \cite{Van88}. It is also
observed in \cite{Mit74} that the EUV emission flare does lead
(but only slightly) the X-ray flare.

\section{Ionospheric response to the solar flare of December 28,
1999}
\label{GLOB-Sect6}

A powerful impulsive flare of December 29, 1999 was chosen also to
illustrate the practical implementation of the proposed method. The
dashed line in Fig. 3d shows the time dependence of the X-ray
emission intensity $R(t)$ of this flare as measured by the
X-telescope BATSE on satellite CGRO in 25--50 keV range (thin line).
The envelope of the dependence $R(t)$ represents a bell-shaped
pulse of about 1.5-min duration with a maximum corresponding to
00:44 UT, or to about 16:00 LT in the U.S. West (Fig. 1b). This time
interval is characterized by a low level of geomagnetic disturbance
(within - 22 nT), which simplified greatly the SID detection problem.
The December 28, 1999 flare was also recorded by X-ray telescope
HXT on the YOHKOH satellite (Fig.~3d - thick line; in 23-33 keV
range).

Fig.~3a presents typical time dependencies of an "vertical" TEC
$I(t)$ for the PRN26 satellite at the LEEP station
($39.1^\circ$N; $241.7^\circ$E) on December 28, 1999 (thick line)
and for PRN04 at the CHI1 station ($60.2^\circ$N; $213.3^\circ$E
-- thin line). It is apparent from Fig.~3a that in the presence of
slow TEC variations, the SID-induced short-lasting sudden
increase in TEC is clearly distinguished in the form of a "step"
as large as 0.2--0.5 $TECU$.

For the same series, similar lines in panel b. show variations of
the time derivative of TEC $dI(t)/dt$ with the linear trend
removed and with a smoothing with the 5-min time window.

The (normalized to $N$) result of a coherent summation of the
$S(t)$-series for all beams and GPS stations located mainly on
the dayside is presented in panel c (thick line).

A comparison of the coherent sum of $S(t)$ with the time
dependence of the X-ray emission intensity $R(t)$ of the
December 28, 1999 flare, based on the data from the X-telescopes
BATSE on satellite CGRO data in 25--50 keV range and HXT on the
YOHKOH satellite in 14-23 keV range (panel d), shows their high
correlation and an almost total coincidence of the form of $S(t)$
with the $R(t)$ pulse envelope.

\section{Discussion and conclusions }
\label{GLOB-Sect7}

This paper has offered a brief outline of the concept of a global
detector (GLOBDET) of atmospheric disturbances, based on phase
measurements of a TEC in the ionosphere which were made using the
international network of two-frequency multichannel GPS
receivers. We have discussed the possibilities of using GLOBDET
method in detecting ionospheric effects of solar flares. A case
study of a global ionospheric response to a powerful solar flares
of July 29, 1999 and December 28, 1999 illustrates the new
experimental potential.

The GLOBDET technology, suggested in this paper, can be used to
detect small solar flares; the body of data processed is the only
limitation in this case. The high sensitivity of GLOBDET permits
us to propose the problem of detecting, in the flare X-ray and
EUV ranges, emissions of nonsolar origins which are the result of
supernova explosions.

For powerful solar flares like the one examined in this report,
it is not necessary to invoke a coherent summation, and the SID
response can be investigated for each beam. This opens the way to
a detailed study of the SID dependence on a great variety of
parameters (latitude, longitude, solar zenith angle, spectral
characteristics of the emission flux, etc.). With current
increasing solar activity, such studies become highly
challenging. In adidtion to solving traditional problems of
estimating parameters of ionization processes in the ionosphere
and problems of reconstructing emission parameters \cite{Mit74},
the data obtained through the use of GLOBDET can be used to
estimate the spatial inhomogeneity of emission fluxes at scales
of the Earth's radius.

The GLOBDET technology can also be used in detecting disturbances
of natural and technogenic origins accompanied by the propagation of
acoustic and acoustic-gravity waves. They include such phenomena as
explosions, rocket launches, solar eclipses, the displacement of the
solar terminator, earthquakes, volcanic eruptions, heavy
thunderstorms, and auroral heating. Unlike solar flares, however, the
response to the above-mentioned effects is not time-coincident for
different beams. Therefore, when selecting complex factors $\tilde
K_i$, it is necessary to appropriately take into account the amplitude
attenuation and the phase delay which are caused by the propagation
decay and finite velocity (sonic and subsonic in this case) of the
response in the ionosphere. Furthermore, a processing of traveling TEC
disturbances implies essentially using regional rather than global GPS
networks, which reduces the sample statistic of the beams when the
signal is accumulated.

\section{Acknowledgments}
Author is grateful to Altyntsev ~A.~T. and ~L.~A. Leonovich for
their encouraging interest in this study and active
participations in discussions. Author is also indebted to
V.~V.~Grechnev, ~E.~A.~Kosogorov, ~O.~S.~Lesuta and ~K.~S.
Palamarchouk for preparing the input data. Thanks are also due
V.~G.~Mikhalkovsky for his assistance in preparing the English
version of the \TeX-manuscript. This work was done with support
from the Russian Foundation for Basic Research (grant
No. 99-05-64753) and from RFBR grant of leading scientific schools of
the Russian Federation No. 00-15-98509.

{}

\begin{thebibliography}{}

\bibitem{Afr98}
Afraimovich, ~E.~L., ~K.~S. Palamartchouk, and ~N.~P. Perevalova,
GPS radio interferometry of travelling ionospheric disturbances,
{\it J. Atmos. and Solar-Terr. Phys., } {\bf 60, } 1205--1223, 1998.

\bibitem{Afr2000}
Afraimovich ~E.~L., ~E.~A. Kosogorov, ~L.~A. Leonovich,~K.~S.
Palamarchouk, ~N.~P. Perevalova, and ~O.~M. Pirog,
Determining parameters of large-scale traveling ionospheric
disturbances of auroral origin using GPS-arrays,
{\it J. Atmos. and Solar-Terr. Phys.} {\bf 61, } 2000 (accepted).

\bibitem{Cal95}
Calais~E. and ~J.~B. Minster,
GPS detection of ionospheric
perturbations following the January 1994, Northridge earthquake,
{\it Geophys. Res. Lett.,} {\bf22, } 1045--1048, 1995.

\bibitem{Cal96}
Calais ~E. and ~J.~B.Minster,
GPS detection of ionospheric
perturbations following a Space Shuttle ascent,
{\it Geophys. Res. Lett.,} {\bf23, } 1897--1900, 1996.

\bibitem{Cal98}
Calais~E.,~M.~A.~H. Hedlin,~M.~A. Hofton, and ~B.~J. Minster,
Ionospheric signature of surface mine blasts from Global Positioning
System measurements,
{\it Geophys. J. Int.,} {\bf132, } 191--202, 1998.

\bibitem{Col85}
Collin~R.~E.
{\it Antennas and Radiowave Propagation,}
McGraw-Hill, New York, 1985.

\bibitem{Dav69}
Davies ~K.
{\it Ionospheric  radio  waves,}
Blaisdell  Publishing Company,  A   Division   of   Ginn   and
Company, Waltham, Massachusetts-Totonto-London, 1969.

\bibitem{Don69}
Donnelly~R.~F,
Contribution of X-ray and EUV bursts of solar flares to sudden
frequency deviations,
{\it J. Geophys. Res.,} {\bf 74, } 1873--1877, 1969.

\bibitem{Fit97}
Fitzgerald~T.~J.,
Observations of total electron content perturbations
on GPS signals caused by a ground level explosion,
{\it J. Atmos. and Solar-Terr. Phys.,} {\bf 59, } 829--834, 1997.

\bibitem{Gur93}
Gurtner,~W.,
RINEX: The Receiver Independent Exchange
Format Version 2.
http://igscb.jpl.nasa.gov:80/igscb/data/format/rinex2.txt, 1993.

\bibitem{Hof92}
Hofmann-Wellenhof,~B., ~H. Lichtenegger, and ~J. Collins,
{\it Global Positioning System: Theory and Practice,}
Springer-Verlag Wien, New York, 1992.

\bibitem{Klo86}
Klobuchar,~J.~A., Ionospheric time-delay algorithm for
single-frequency GPS users,
{\it IEEE Transactions on Aerospace and Electronics System,}
{\bf AES 23(3), } 325--331, 1986.

\bibitem{Klo97}
Klobuchar,~J.~A., Real-time ionospheric science: The new
reality,
{\it Radio Science,} {\bf 32, } 1943--1952, 1997.

\bibitem{Man98}
Mannucci,~A.~J., ~C.~M. Ho, ~U.~J. Lindqwister, ~T.~F. Runge,
~B.~D. Wilson and ~D.~N. Yuan, A global mapping technique
for GPS-drived ionospheric TEC measurements,
{\it Radio Science,} {\bf 33, } 565--582, 1998.

\bibitem{Men74}
Mendillo~M.,~J.~A. Klobuchar,~R.~B. Fritz,~A.~V.~da~Rosa,~L.
Kersley, ~K.~C. Yeh,~B.~J. Flaherty,~S. Rangaswamy,~P.~E.
Schmid,~J.~V. Evans, ~J.~P. Schodel,~D.~A. Matsoukas,~J.~R.
Koster,~A.~R. Webster,~P. Chin,
Behavior of the Ionospheric F Region During the Great Solar Flare of
August 7, 1972,
{\it J. Geophys. Res.} {\bf 79, } 665--672, 1074.

\bibitem{Mit74}
Mitra~A.~P.,
{\it Ionospheric effects of solar flares,}
New Delhi -12, India, 1974.

\bibitem{Sai98}
Saito~A., ~S. Fukao and ~S. Miyazaki,
High resolution mapping of TEC perturbations with the GSI GPS
network over Japan,
{\it Geophys. Res. Lett.,} {\bf25, } 3079--3082, 1998.


\bibitem{Tho71}
Thome~G.~D and ~L.~S.Wagner,
Electron density enhancements in the E and F regions of the
ionosphere during solar flares,
{\it J. Geophys. Res.,} {\bf 76, } 6883--6895, 1971.

\bibitem{Van88}
Vanderveen~K., L.~E.~Orwig, and E.~Tandberg-Hanssen,
Temporal correlations between impulsive ultraviolet and hard X-ray
busts in solar flares observed with high time resolution,
{\it Astrophys. J.,} {\bf 330, } 480--492, 1988.
\end{thebibliography}
\end{document}